\documentclass[10pt, conference, letterpaper]{IEEEtran}
\IEEEoverridecommandlockouts
\usepackage{cite}
\usepackage{amsmath,amssymb,amsfonts}
\usepackage{algorithmic}
\usepackage{graphicx}
\usepackage{textcomp}
\usepackage[dvipsnames]{xcolor}
\usepackage{alphabeta}
\usepackage{xspace}
\usepackage{color}
\usepackage[caption=false]{subfig}
\usepackage{makecell}
\usepackage{algorithm}
\usepackage{array}
\usepackage{stfloats}
\usepackage{url}
\usepackage{verbatim}
\usepackage[normalem]{ulem}
\usepackage{mathtools}
\usepackage{enumitem}
\usepackage{titlesec}

\titlespacing*{\subsection}{0pt}{2pt}{2pt}

\usepackage{cuted}

\newtheorem{remark}{Remark}
\newlist{Properties}{enumerate}{2}
\setlist[Properties]{label=Property \arabic*., font=\textbf, itemindent=*}

\DeclareRobustCommand{\erase}{\bgroup\markoverwith{\textcolor{red}{\rule[.5ex]{2pt}{0.5pt}}}\ULon}

\def\BibTeX{{\rm B\kern-.05em{\sc i\kern-.025em b}\kern-.08em
    T\kern-.1667em\lower.7ex\hbox{E}\kern-.125emX}}

\begin{document}
\title{A Novel One-tap Equalizer for Zero-Padded AFDM System over Doubly Selective
Channels

%A Novel Zero-Padded AFDM Scheme with One-tap Frequency-of-Affine Domain Equalizer over Doubly Selective Channels
\vspace*{-4mm}}

%\title{Enabling One-Tap Equalizer for Doubly Selective Channels: Zero-Padded AFDM with Frequency of Affine Domain Equalizer}

\author{\IEEEauthorblockN{Chenyang Zhang, Akram Shafie, Cheng Shen, Deepak Mishra, and Jinhong Yuan}\\
\vspace*{-3mm}
\IEEEauthorblockA{\
School of Electrical Engineering and Telecommunications, University of New South Wales, Sydney, NSW 2052,  Australia \\\vspace*{-4mm}
Emails:\{chenyang.zhang1, akram.shafie, cheng.shen1, d.mishra, j.yuan\}@unsw.edu.au}\vspace*{-4mm}
}
\maketitle
\begin{abstract}
Recently, affine frequency division multiplexing (AFDM) has gained traction as a robust solution for doubly selective channels. In this paper, we present a novel low-complexity one-tap equalizer for zero-padded AFDM (ZP-AFDM) systems. We first select the AFDM parameters, $c_1$ and $c_2$, such that $c_1$ has a relatively high value, and $c_2$ depends on $c_1$, which simplifies the affine domain input-output relation (IOR).  This selection also demonstrates that a phase term that varies slowly along the affine domain is experienced by all affine domain symbols and this variation is significantly slower compared to that experienced by the time domain symbols over doubly selective channels. To simplify the equalization, we then introduce zero padding to the transmitted affine domain symbols  and reconstruction operation on the received affine domain symbols. By doing so, we convert the effective affine domain IOR of our ZP-AFDM system to be characterized using approximately circular convolution. Next, we transform the resulting affine domain symbols into a new domain called the frequency-of-affine (FoA) domain. We propose our one-tap equalizer in this FoA domain to efficiently recover the transmitted symbols. Numerical results demonstrate the effectiveness of our proposed one-tap equalizer, particularly when $c_1$ is high, without compromising performance robustness.
\end{abstract}

% \begin{IEEEkeywords}
% Affine frequency division multiplexing, one-tap equalizer, chirp modulation, linear time-varying channels, high mobility communications.
% \end{IEEEkeywords}
\vspace{-2mm}

\section{Introduction}
%\IEEEPARstart{O}{rthogonal} 
%Orthogonal frequency division multiplexing (OFDM) modulation has been at the forefront of 4G, 5G, Wi-Fi and other wireless communication systems for the last couple of decades, thanks to its simple transceiver implementation and the utilization of one-tap equalization at the receiver for linear time-invariant (LTI) channels \cite{andrews2014will}. 
%However, it remains unclear whether OFDM can support reliable communication in some of the future-envisioned communication scenarios characterized by doubly selective linear time-variant (LTV) channels, such as high mobility and underwater acoustic \cite{2021_WCM_JH_OTFS}. 
%The severe Doppler spread in these scenarios compromises the subcarrier orthogonality of OFDM. 
%This requires more sophisticated receivers or looking beyond OFDM for new modulation schemes to enhance communication system resilience in LTV channels.

Orthogonal frequency division multiplexing (OFDM) has been a key technology in 4G, 5G, Wi-Fi, and other wireless systems due to its simple transceiver implementation and one-tap equalization at the receiver for linear time-invariant (LTI) channels \cite{andrews2014will}. However, it remains unclear whether OFDM can support reliable communication in some of the future-envisioned communication scenarios with doubly selective linear time-variant (LTV) channels, such as high mobility and underwater acoustic \cite{2021_WCM_JH_OTFS}, where severe Doppler spread compromises subcarrier orthogonality. This requires more sophisticated receivers or new modulation schemes to enhance communication system resilience in doubly selective channels.

%With the rapid advancement of novel future applications especially in high-mobility environments, the sixth-generation (6G) wireless communication system has attracted significant attention, particularly in addressing communication challenges in high-mobility environments such as unmanned aerial vehicles (UAVs), high-speed railways, autonomous vehicles, and non-terrestrial networks. These scenarios pose large Doppler shifts, which can degrade performances of classical modulation technique signals, especially for that of orthogonal frequency division multiplexing (OFDM) modulation.

To combat Doppler effects and improve the resilience of communication systems, various multicarrier modulation techniques have been recently introduced and studied \cite{rou2024otfs}. A leading approach, delay-Doppler (DD) modulation, including  orthogonal time frequency space (OTFS) modulation \cite{hadani2017orthogonal} and orthogonal DD division multiplexing
(ODDM) \cite{lin2022orthogonal}, operates within the DD domain, couples the information symbols with the channel representation in the DD domain, and offers robust performance against channel
fading and time variations that adversely affect OFDM systems. 
%, and offers robust performance against channel fading and time variations that adversely affect OFDM systems. OTFS and ODDM modulations provide diversity across both time and frequency domains, making it well-suited for high-mobility scenarios. % \cite{surabhi2019diversity,lin2022orthogonal}. 

As an alternative to OTFS and ODDM, affine frequency division multiplexing (AFDM) modulation was proposed by Bemani \textit{et al.} \cite{bemani2021afdm}. 
It multiplexes information symbols within the affine domain\footnote{In this paper, we use the term `affine' to denote the domain in which AFDM symbols are embedded.}, which utilizes the inverse of discrete affine Fourier transform (DAFT) to obtain the discrete time domain symbols.  %The AFDM modulation can be viewed as if it carries each information symbol using a sequence of chirps.
AFDM achieves effective path separation in the affine domain and equal gain across data symbols, enabling BER performance on par with OTFS and ODDM. % in terms of BER. 
%AFDM’s unique approach in the affine domain has shown promise, especially under conditions that would otherwise compromise conventional techniques like OTFS and ODDM \cite{haif2024novel}.

%To enhance the efficiency of detection in OTFS and AFDM systems, various algorithms have been developed. 
Various equalization algorithms have been developed for AFDM, primarily focusing on affine domain approaches.
Bemani \textit{et al.} \cite{bemani2022low} presented a low-complexity decision feedback equalizer (DFE) based on weighted maximal ratio combining (MRC), designed to perform effectively while retaining lower complexity than linear minimum mean square error (LMMSE) alternatives.
Wu \textit{et al.} proposed a message-passing algorithm (MPA) for AFDM, resulting in improved performance compared to traditional MMSE and MRC methods \cite{wu2024afdm}. 
Similar to that proposed for OTFS and ODDM, other iterative equalizations and multi-tap equalizations focusing on time and/or frequency domain symbols can also be utilized \cite{OTFS_MMSE}. 
Despite the advantages of these equalization algorithms, they mostly remain computationally intensive, especially as compared to the single-tap equalization adopted in OFDM systems for linear-time invariant channels.

In this paper, we propose a low-complexity one-tap equalizer specifically designed for the zero-padded AFDM (ZP-AFDM) systems over doubly selective channels. 
To make this one-tap equalizer possible, we first strategically select AFDM post-chirp parameter $c_1$ and pre-chirp parameter $c_2$ such that they satisfy $4c_1 c_2 N^2=1$. By doing so, we establish a simplified input-output relation (IOR) for the AFDM system. This IOR reveals that affine domain symbols experience two phase term. 
One phase term, which varies slowly along the affine domain, is experienced by all affine domain symbols and this variation is significantly slower compared to that experienced by the time domain symbols over doubly selective channels, while the other is called the additional phase term, which only affects the symbols at the beginning and the ending of received affine domain symbol vector.
%
%
%One is called the low-frequency phase term, which varies significantly slowly over the affine domain symbols as compared to that experienced by time domain symbols, while the other is called the additional high-frequency phase term, which only affects the symbols at the beginning and the ending of affine domain symbols.
%
Next, we introduce zero padding to the transmitted affine domain symbols and reconstruction operation in the affine domain of the receiver. This modification enables to overcome the impact of the additional phase term and to transform the affine domain IOR of ZP-AFDM to be characterized by a approximate circular convolution with a slowly varying phase term. 

We then transform the reconstructed affine domain symbols into what we call the frequency-of-affine (FoA) domain using Fourier transform.
Within this domain, we show that our proposed one-tap equalizer can be effectively adopted to recover the transmitted signal.
Using numerical results, we show the effectiveness of our proposed equalizer, especially when $c_1$ is high. 
For example, with $10\%$ overhead, the proposed scheme has no error floor up to BER of $1\times10^{-6}$. This demonstrates
the effectiveness of our transmission scheme with dramatically reduced receiver complexity.

%We highlight that our proposed one-tap equalization for ZP-AFDM significantly simplifies the equalization process without compromising robustness performance,
%thereby addressing the key bottleneck of high equalization complexity faced by recently proposed OTFS, ODDM, and AFDM modulation designed for doubly selective channels.
\section{Classical AFDM System Model}
\begin{figure*}
   % \vspace*{-0.6cm} 
    \centering
        \centering
         \includegraphics[width=0.87\linewidth]{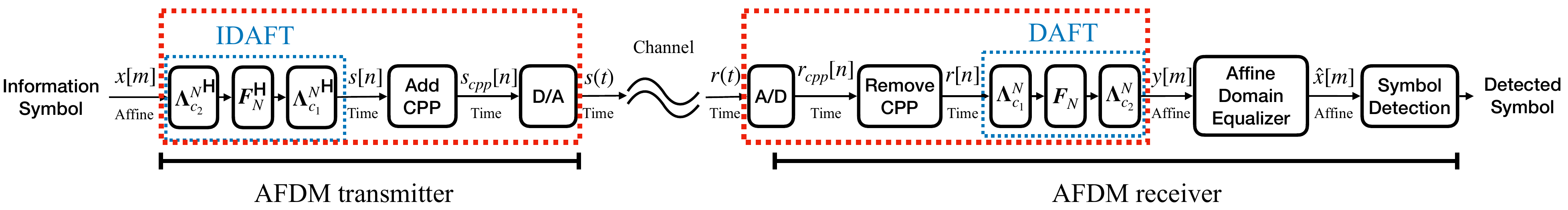}
         \vspace*{-0.2cm} 
         \caption{Classical AFDM System Model.}
         \label{fig:AFDM}
         \vspace*{-0.7cm} 
\end{figure*}

This section illustrates the classical AFDM system model, whose schematic representation is described in Fig. \ref{fig:AFDM}.

\subsection{Transmitter}

~~Suppose $N$ information-bearing symbols are transmitted within the allocated bandwidth $B=\frac{1}{T_\textrm{s}}$ Hz, and the transmission time $T=N T_\textrm{s}$ \textit{sec}, where $T_\textrm{s}$ represents the sampling period. 
In AFDM, let us denote these $N$ information-bearing symbols in the affine domain by $x[m], \forall m\in\{0,1,\cdots, N-1\}$, where $m$ is the symbol index in the affine domain. 
%
%The choice of  $N_\textrm{d}$ and $N_z$ and the order in which the information-bearing symbols and zero-padded symbols will be analyzed in the later in section .
% Here, the information-bearing symbols are located at the certain transmitted affine domain symbols, expressed as
% \begin{equation}
%     x[m] = \begin{cases}
%         x_d[m],&k_\textrm{max}<m<N-c_1*l_max-k_max\\
%         0,&\textrm{otherwise}
%     \end{cases},
% \end{equation}
% where $x_d[m]$ represents the information-bearing symbols.
%
The transmitter processing commences with an IDAFT applied to $x[m]$ to yield the discrete-time domain symbols $s[n], \forall n\in\{0,1,\cdots, N-1\}$, given by
\begin{equation}
\begin{aligned}
s[n] = &\frac{1}{\sqrt{N}}\sum_{m=0}^{N-1}x[m]e^{j 2 \pi \left( c_1 n^2+c_2 m^2+ \frac{mn}{N} \right)},
\end{aligned}
\label{equ:s}
\end{equation}
where $c_1$ and $c_2$ are the post-chirp and pre-chirp parameters used in IDAFT/DAFT, respectively \cite{Shen2025TWC}.
%
% It can be further expressed as 
% \begin{equation}
% \begin{aligned}
%     A(m,n)  = \frac{1}{\sqrt{N}}&\underbrace{e^{-j 2 \pi c_1 n^2}}_\text{Term 1}
%     \underbrace{\frac{1}{\sqrt{N}}e^{-j 2 \pi \frac{mn}{N}}}_\text{Term 2}
%     \underbrace{e^{-j 2 \pi c_2 m^2}}_\text{Term 3}.
%     \label{equ:Amn}
% \end{aligned}
% \end{equation}
% When carefully observing the decomposition of $A(m,n)$ in \eqref{equ:Amn}, Term 2 corresponds to the DFT, while Term 1 and Term 3 account for additional quadratic phases that depend on the c1 and c2, respectively.

%
% The discrete-time-domain symbols $s[n]$ can be further expressed as 
% \begin{equation}
%     s[n] = \frac{1}{\sqrt{N}}\sum_{m=0}^{N-1} x'[m] \Psi_{c_m}[n],
%     \label{equ:x'm}
% \end{equation}
% where $x'[m] = x[m]e^{j 2 \pi c_2 m^2}$ and $\Psi_{c_m}[n] = e^{j 2 \pi (\frac{mn}{N}+c_1 n^2)}$.
% It can be seen that from \eqref{equ:s} that $x[m]$ is firstly given a constant phase shift $e^{j 2 \pi c_2 m^2}$, and then is carried by a linear chirp sequence with a unified chirp rate $c_1$ and differential frequency shift $e^{j 2 \pi \frac{mn}{N} }$\footnote{The discrete Fourier transform (DFT) used in OFDM, OTFS and ODDM and the discrete Fresnel transform (DFnT) used in OCDM are two distinct special cases of DAFT. The DAFT becomes DFT when $c_1 = c_2 = 0$, while DAFT becomes DFnT when $c_1 = c_2 = \frac{1}{2N}$.}. The parameters $c_1$ and $c_2$ will be discussed in detail in later sections. 
%It can be compared with OCDM \cite{ouyang2016orthogonal}.

%figure shows instantaneous frequency 

Next, a chirp periodic prefix (CPP) with length $L_c\geq l_\textrm{max}$ is introduced %for the symbol $s[n], \forall n\in\{0,1,\cdots, N-1\}$. This is to avoid ISI and to preserve the continuity of each chirp between the prefix and the symbols $s[n]$. \color{black} %to obtain $\boldsymbol{s}_\textrm{cpp}$. 
%
%This 
to the time domain symbols to yield % the CPP-added symbols as %$s_{\textrm{cpp}}[m]$% becomes
%\footnote{\color{red} When $2Nc_1$ is an integer value and $N$ is even, CPP effectively becomes equivalent to the CP.\color{black}}
\begin{align}
\begin{split}
    \!\!\!\!\!\!s_{\textrm{cpp}}[n]=&\begin{cases}
        s[n+N]e^{-j2\pi c_1\left(N^2+2Nn\right)}, & -L_c\leq n\leq-1,\\
        s[n], &0\leq n\leq N-1.
    \end{cases}
    \label{equ:scpp1}
\end{split}
% \begin{split}
%     =&\frac{1}{\sqrt{N}}\sum_{m=0}^{N-1}x[m]e^{j 2 \pi \left( c_1 n^2+c_2 m^2+ \frac{mn}{N} \right)}.
% \label{equ:scpp2}
% \end{split}
\end{align}
As can be observed in~\eqref{equ:scpp1}, CPP is essentially the same as the CP used in conventional multicarrier modulation schemes like OFDM, OTFS, ODDM, etc., but with an additional phase term $e^{-j2\pi c_1(N^2+2Nn)}$ \cite{bemani2023affine}. CPP in AFDM preserves the continuity of each chirp between the prefix and
the symbols $s[n]$, and avoids interference between consecutive AFDM frames.

Finally, $s_{\textrm{cpp}}[n]$, $\forall n\in\{-L_c,\cdots, N-1\}$, are sent through a digital-to-analog (D/A) converter parameterized by the pulse shaping filter $a(t)$ and the sampling period $T_s {=} \frac{1}{B}$ to obtain the transmit-ready signal % as
\begin{align} 
\label{equ:st}
    s(t)&=\sum_{n=-L_{c}}^{N-1}s_{\textrm{cpp}}[n]a(t-n T_s ).
\end{align}

\subsection{Doubly-Selective Channel}
~~In the high-mobility scenarios, the transmitted signal $s(t)$ propagates through a doubly selective channel, which is modelled via $P$ significant propagation paths. 
The $i$-th resolvable path, where $i \in \{1,\dots,P\}$, 
is described by the corresponding complex channel coefficient $h_i$, path delay $\tau_i \in [0,\tau_{\textrm{max}}]$, and Doppler shift $\upsilon_i \in [-\upsilon_{\textrm{max}},\upsilon_{\textrm{max}}]$, where $\tau_{\textrm{max}}$ and $\upsilon_{\textrm{max}}$ are the maximum delay and Doppler shift, respectively.
Under such considerations, the doubly selective channel can be represented based on the time-variant impulse response function as
    $g(t,\tau) = \sum_{i=1}^{P} h_i e^{j 2\pi \upsilon_i t} \delta(\tau-\tau_i).
    \label{equ:g_t}$
The received signal can be written as
\begin{equation}
\begin{aligned}
    r(t) %= &\int_{0}^{\tau_{\textrm{max}}} s(t-\tau) g (t,\tau) d\tau + w(t)\\
    =&\sum_{i=1}^{P} h_i s(t-\tau_i)e^{j 2\pi \upsilon_i t}+ w(t),
\end{aligned}
\end{equation}
where $w(t)\thicksim \mathcal{CN}(0,\sigma^2)$ is the complex additive white Gaussian noise and $\sigma^2$ is the noise variance.

\begin{figure*}[b]
\vspace*{-0.6cm} 
\hrulefill
\vspace*{-0.05cm} 
\begin{align}
    y[m] %=&\sum_{i=1}^{P}\sum_{m'=0}^{N-1}h_i x[m'] e^{\frac{j2\pi n}{N}\left( c_1 N l_i^2-m l_i+ c_2Nm'^2-c_2 N m^2\right)}|_{m'=\left(m-k_i+2c_1Nl_i\right)_N}+\bar{w}[m]\notag \\
    =&\sum_{i=1}^{P} h_i x[(m-k_i+2c_1Nl_i)_N] D_i[m]  e^{\frac{j2\pi m}{N}\left((4c_1c_2N^2-1)l_i-2c_2 N k_i\right)} e^{\frac{j2\pi}{N}\left((4c_1c_2N^2-1) (c_1Nl_i^2-l_i k_i)+c_2N k_i^2\right)}+\bar{w}[m]. \tag{8} \label{equ:chirpIO}
\end{align}
\vspace*{-0.6cm} 
\end{figure*}
\subsection{Receiver}
~~At the receiver, $r(t)$ is passed through a matched filter and an analog-to-digital (A/D) converter %parameterized by $a(t)$ and sampling period $T_s$ 
to obtain the received time domain sampled symbols %\vspace*{-2mm}
\begin{equation}
    \begin{aligned}
        r_\textrm{cpp}[n] = &\sum_{i=1}^{P} h_i s_{\textrm{cpp}}[n-l_i]e^{j2\pi k_i\frac{n}{N}}+ w[n],
    \end{aligned}
    \label{equ:rcpp}
\end{equation}
for $ n\in\{-L_c,\cdots, N-1\}$, where $l_i = \frac{\tau_i}{T_s}$ and $k_i = \upsilon_i T_s N$ are the normalized delay and Doppler of $i$-th path, respectively, 
$w[n]\thicksim \mathcal{CN}(0, \sigma^2)$ is complex noise, and $l_\textrm{max} = \frac{\tau_\textrm{max}}{T_s}$ and $k_\textrm{max} = \upsilon_\textrm{max} T_s N$ are the maximum normalized delays and Doppler shifts, respectively.\footnote{Due to the limited time and frequency domain resources, $l_i$ and $k_i$ may not necessarily be on-grid w.r.t. the delay and Doppler resolution $T_s$ and $\frac{1}{NT_s}$, respectively. This phenomenon is commonly referred to as the off-grid/fractional delay and Doppler shift \cite{bemani2021afdm}. However, following \cite{lin2022orthogonal,hadani2017orthogonal, bemani2022low}, for simplicity in presentation, 
we assume $l_i$ and $k_i$ to be on-grid, meaning $l_i$ and $k_i$ are integers. Off-grid scenario will be discussed in the journal. 
%Despite this simplification, when $l_i$ and/or $k_i$ are off-grid, the contributions from each channel path can be approximated as the combination of multiple paths with neighboring on-grid paths\cite{tong2024orthogonal}. 
%In this case, \eqref{equ:rcpp} is still valid, in the sense, it represents the sampled equivalent channels \cite{tong2024orthogonal}. 
}
Then the CPP is removed to obtain %the received time domain symbols as 
\begin{align}
    r[n] = r_\textrm{cpp}[n], ~~~~~~0\leq n\leq N-1.
    \label{equ:r}
\end{align}
% \begin{equation}
%     \begin{aligned}
%         r[n] = &r_\textrm{cpp}[n], ~~~~~~~~0\leq n\leq N-1 \\
%         = &\sum_{i=1}^{P} h_i s_{\textrm{cpp}}[n-l_i]e^{\frac{j 2\pi}{N} k_i n}+ w[n]. %, 0\leq n\leq N-1.
%     \end{aligned}
%     \label{equ:r}
% \end{equation}
Thereafter, DAFT is applied to $r[n] $ for $n\in\{0,1,\cdots, N{-}1\}$ to obtain the received affine domain symbols %$y[m]$ for $m\in\{0,1,\cdots, N{-}1\}$ as
\begin{equation}
\begin{aligned}
    y[m]&=\frac{1}{\sqrt{N}}\sum_{n=0}^{N-1}r[n]e^{-j 2 \pi \left( c_1 n^2+c_2 m^2+ \frac{mn}{N} \right)}, 
\end{aligned}
\label{equ:y}
\end{equation}
for $m\in\{0,1,\cdots, N{-}1\}$. Finally, equalization is performed on $y[m]$ to recover the transmitted symbols $x[m]$.

\subsection{AFDM Affine Domain Input-Output Relation (IOR)}
~~By substituting \eqref{equ:r} in \eqref{equ:y}, the affine domain  IOR for AFDM can be obtained as in \eqref{equ:chirpIO} \cite{bemani2021afdm}.%, shown at the bottom of next page 
Therein, $D_i[m],\forall i\in \{1,\dots,P\},m\in\{0,\dots,N-1\}$ represents an additional phase term introduced by the $i$-th channel path to the $m$-th received symbol and $\bar{w}[m]$ is affine domain noise samples. $D_i[m]$ is expressed for $k_i-2c_1Nl_i<0$ as 
\begin{align}\setcounter{equation}{8}
    &D_i[m] \ \\
    &=\begin{cases}
        e^{j2\pi c_2 N \left( N-2(m-k_i +2 c_1 N l_i)\right)}, &m\geq N+k_i-2c_1Nl_i,\notag\\
        1, &\textrm{otherwise},
    \end{cases}
\end{align}
and for $k_i-2c_1Nl_i\geq 0$ as
\begin{align}
    \!\!\!D_i[m]=\begin{cases}
        e^{j2\pi c_2 N \left( N+2(m-k_i +2 c_1 N l_i)\right)}, &m< k_i+2c_1Nl_i,\\
        1, &\textrm{otherwise}.\\
    \end{cases}%\notag
\end{align}

As can be observed in \eqref{equ:chirpIO}, the affine domain IOR is a combination of several transmitted symbols, necessitating a multitap equalizer. In the literature, MRC and MPA are used for AFDM equalizers, but their computational complexity is extremely high \cite{bemani2022low,wu2024afdm}. 
To overcome this challenge, in the next section, we propose an AFDM transmission scheme with a simplified one-tap equalizer. % and simplified equalizer for the proposed AFDM system.

\section{Proposed Zero-Padded AFDM System with One-Tap Equalizer}

\begin{figure*}[t]
    %\vspace*{-0.2cm} 
    \centering
        \centering
         \includegraphics[width=0.95\linewidth]{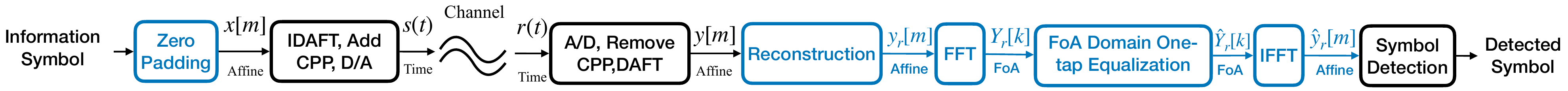}
         \vspace*{-0.2cm} 
         \caption{Proposed zero-padded AFDM system with one-tap equalizer in FoA domain.}
         \vspace*{-0.4cm} 
         \label{fig:AFDMonetap}

\end{figure*}

The schematic representation of our proposed novel zero-padded AFDM (ZP-AFDM) scheme with FoA domain one-tap equalizer is shown in Fig. \ref{fig:AFDMonetap}, 
%In this section, we propose a novel zero-padded AFDM (ZP-AFDM) scheme with FoA domain one-tap equalizer. Its schematic representation is shown in Fig. \ref{fig:AFDMonetap}, 
with the operation marked in blue representing the modified or newly added operation in our proposed system compared to the classical AFDM systems given in Fig. \ref{fig:AFDM}.
We first discuss $c_1$ and $c_2$ parameter selection in our system. %\color{blue}
Next, we discuss the two newly added operations in our proposed system, namely transmitter side affine domain zero-padding %\footnote{%\color{blue}Transmitter side affine domain zero-padding has been employed in \cite{bemani2023affine} for channel estimation and equalization. However, the zero-padding employed in this work serves a distinct purpose compared to those usages.} 
and receiver side cyclically superimposed signal reconstruction. Then, we analyze the FoA domain representation for the modified system. Finally, we propose a novel low-complexity one-tap equalization for the ZP-AFDM scheme.

% In this section, we propose a ZP-AFDM system with a one-tap equalizer, and its schematic representation is shown in Fig. \ref{fig:AFDMonetap}, where the process in the red dotted box in Fig. \ref{fig:AFDM} is shown in one block, respectively. The operation marked in blue represents the modified or newly added operation in our proposed system compared to the classical AFDM systems.
% We first discuss the selection of the two parameters in AFDM $c_1$ and $c_2$. Then, we discussed zero-padding and reconstruction involved in our proposed system. Finally, we introduce the FoA domain one-tap equalization for the ZP-AFDM system.

\subsection{Parameter Selection and Simplified Affine Domain IOR}
~~The DAFT in AFDM is characterized by two important parameters $c_1$ and $c_2$.
As detailed in \cite{bemani2021afdm}, $c_1$ has to be larger than $ \frac{k_\textrm{max}}{N}$ to achieve full
diversity. Considering this lower bound, 
$c_1= \frac{2k_\textrm{max}+1}{2N}$ has been widely used in the literature \cite{bemani2023affine,wu2024afdm}. Different from them, in this work, we propose $c_1$ to be $\chi$ time that of $\frac{2k_\textrm{max}+1}{2N}$ with $\chi>1$. Thus, $c_1=\chi \frac{2k_\textrm{max}+1}{2N}$. The reason for this selection will be detailed in Section III-C.

By closely examining the affine domain IOR in \eqref{equ:chirpIO}, we find that it can be simplified when $4c_1c_2N^2 = 1$.  Moreover, this simplification will enable us to perform simpler equalization that will be discussed in the next section. Considering these, in this work, we consider %$c_2$ to be 
$c_2 = \frac{1}{4c_1 N^2}$, such that it is a function of $c_1$ and $\chi$. 
Under these considerations, affine domain IOR in \eqref{equ:chirpIO} can be further simplified as 
\begin{align}
    y[m] =\sum_{i=1}^{P} h_i &x[(m-k_i+2c_1Nl_i)_N]
    e^{\frac{j\pi k_i^2}{2c_1N^2
    }} \label{equ:affineIO} \\
    ~~~~~~~&\times e^{-j2\pi \frac{k_i}{2c_1N}\frac{m}{N}}  D'_i[m]+\bar{w}[m],\notag
\end{align}
where the new simplified additional phase term %$D'_i[m]$ 
for $k_i-2c_1Nl_i < 0$ becomes
\begin{align}
    \!\!\!\!\!D'_i[m] =\begin{cases}
        e^{j2\pi\left( \frac{1}{4c_1}-\frac{(m-k_i)}{2c_1N}\right)}, &m\geq N+k_i-2c_1Nl_i,\\
        1, &\textrm{otherwise},
    \end{cases}
    \label{equ:Deffective}
\end{align}
and for $k_i-2c_1Nl_i\geq 0$, becomes
\begin{align}
    D'_i[m] =\begin{cases}
        e^{j2\pi\left( \frac{1}{4c_1}+\frac{(m-k_i)}{2c_1N}\right)},&m< k_i-2c_1Nl_i,\\
        1, &\textrm{otherwise}.\\
    \end{cases}
    \label{equ:Deffective2}
\end{align}

\begin{remark}
When closely examining the affine domain IOR in \eqref{equ:affineIO} and comparing it with the time domain IOR in \eqref{equ:rcpp}, we observe several significant similarities between them, along with some differences. 
In particular, as shown in \eqref{equ:rcpp}, a single channel path shifts the time domain symbol by  $\tau_{i,\textrm{time}}=l_i$ and introduces a phase term that varies over the time domain index, characterized by the frequency $f_{i,\textrm{time}}=k_i$. Similar to that, as shown in \eqref{equ:affineIO},  a single channel path shifts the affine domain symbol, but now by $\tau_{i,\textrm{aff}}=k_i - 2c_1Nl_i$. 
Also, it introduces a phase term that varies over the affine domain index $m$ but is now characterized by the effective frequency $\bar{f}_{i,\textrm{aff}}=\frac{k_i}{2c_1N}$.  %$\bar{f}_{i,\textrm{aff}}=\frac{k_i}{2c_1N}=\frac{k_i}{\chi(2k_\textrm{max}+1)}$
Moreover, 1) all the affine domain symbols $y[m]$ experience a $c_1$ and $k_i$-dependent constant phase term $e^{\frac{j\pi k_i^2}{2c_1N^2}}$, and 2) some symbols $y[m]$ experience a $c_1$ and $k_i$-dependent additional phase term $D'_i[m]$.
\end{remark}

From Remark 1, we can observe that $\bar{f}_{i,\textrm{aff}}=\frac{k_i}{2c_1N}=\frac{f_{i,\textrm{time}}}{2c_1N} < f_{i,\textrm{time}}$, indicating that the phase variation/effective frequency experienced by affine domain symbols $y[m]$ is less than that experienced by time domain symbols, Moreover, it is evident that $\left|\bar{f}_{i,\textrm{aff}}\right|<0.5$, $\forall i=\{1,\cdots,P\}$.
Furthermore, it can be observed 
although $c_1=\frac{2k_\textrm{max}+1}{2N}$ has been adopted in previous AFDM studies \cite{bemani2023affine,wu2024afdm}, by increasing $c_1$ such that it is  $c_1=\chi \frac{2k_\textrm{max}+1}{2N}$, where $\chi \in \mathbb{R}$, it is possible to further decrease $\bar{f}_{i,\textrm{aff}}$ significantly.

%A small $\bar{f}_{i,\textrm{aff}}$ and the resulting small phase variation experienced by affine domain symbols $y[m]$  motivates us to explore the potential of transforming  $y[m]$ into a \textit{new domain by using the Fourier transform, which we refer to as the frequency-of-affine (FoA) domain}.  Then, we aim to perform equalization in a low-complexity manner in that FoA domain. 

%Two modifications we introduce to our system to enable the smooth execution of the above operations will be discussed in the next subsection (Section III-B). Finally, in Section III-C, we introduce the proposed FoA domain and one-tap equlizer. 

\subsection{Zero-padding and Cyclically Superimposed  Reconstruction}

\begin{figure*}[t]
\centering
\subfloat[\label{fig:Hc} $\boldsymbol{y}$, $ \boldsymbol{x}$, and $ \boldsymbol{H}_\textrm{aff}$ in \eqref{equ:He} (without zero-padding)]{ 
    \includegraphics[width=0.425\linewidth]{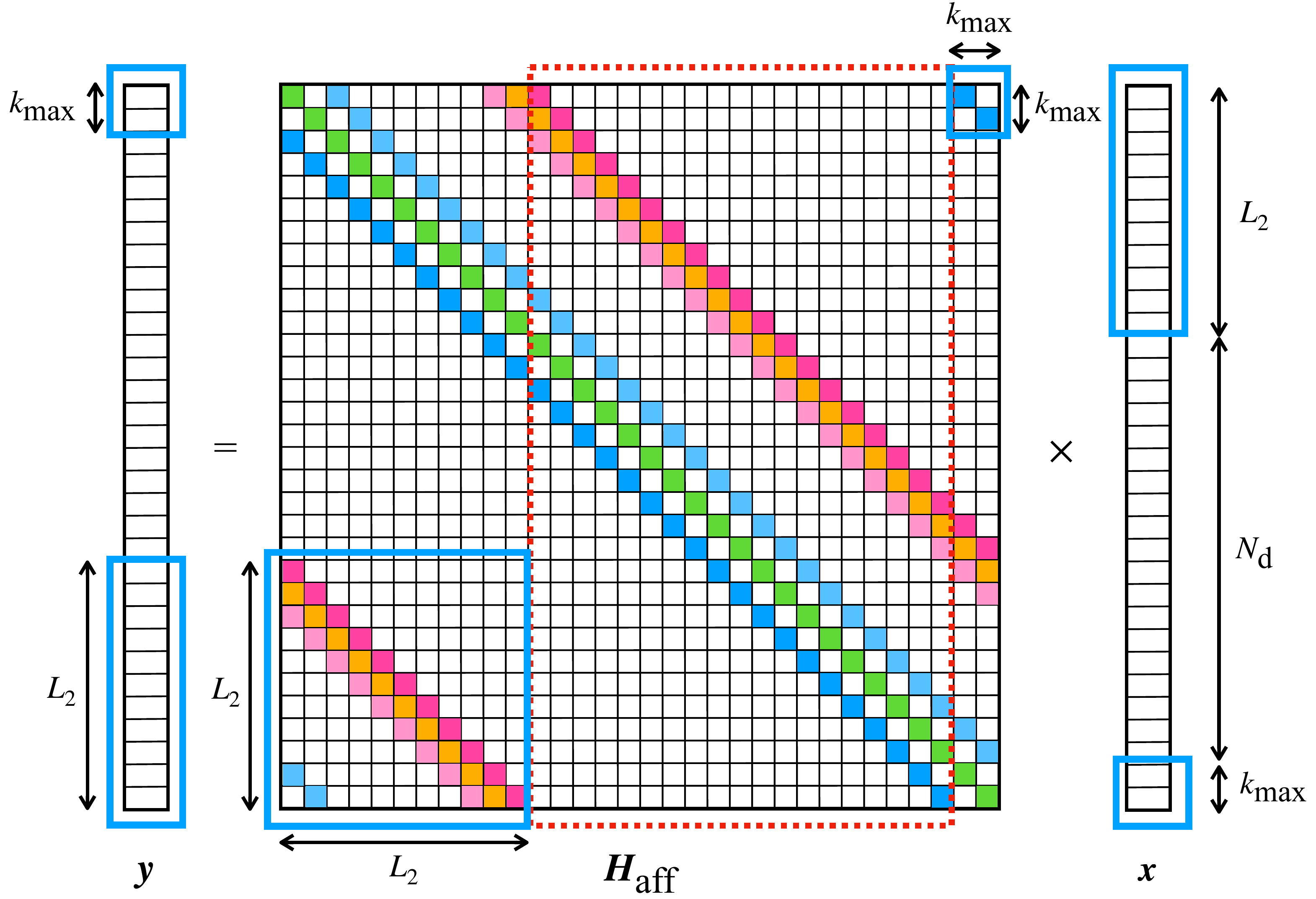}}
    \hfill \hspace{-15mm}
\subfloat[\label{fig:Hcb}$\boldsymbol{y}$, $ \boldsymbol{x}_\textrm{d}$, and $\boldsymbol{\bar{H}}_\textrm{aff}$ in \eqref{equ:He'} %(with zero-padding)
]{ 
    \includegraphics[width=0.285\linewidth]{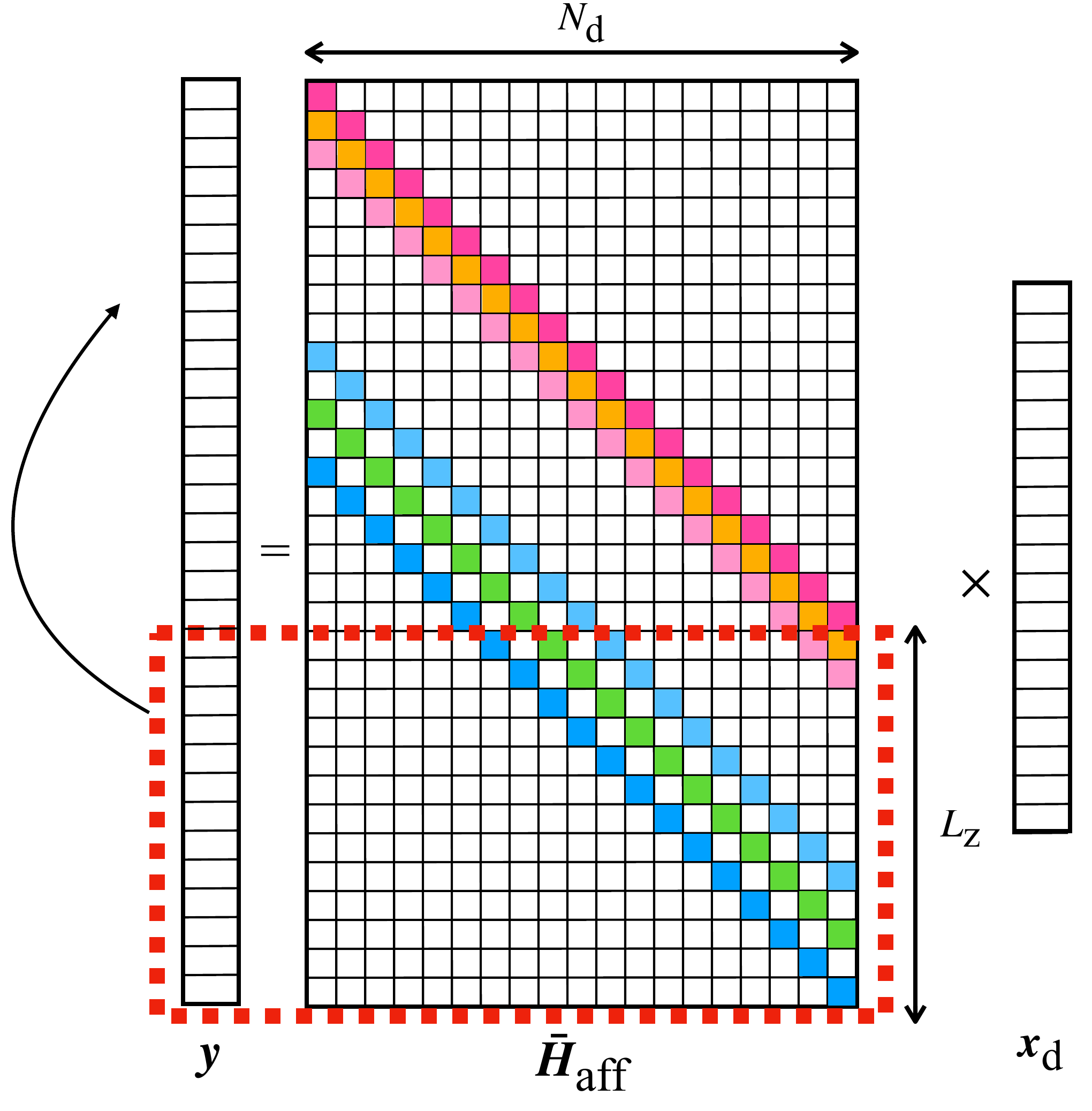}}
    \hfill \hspace{-15mm}
\subfloat[\label{fig:Hcr}$\boldsymbol{y}_\textrm{d}$, $ \boldsymbol{x}_\textrm{d}$, and $\boldsymbol{\hat{H}}_\textrm{aff}$ in \eqref{equ:yrMat} %(with zero-padding and cyclically superimposed signal reconstruction)
]{
    \includegraphics[width=0.21\linewidth,trim={0 -450 0 0},clip]{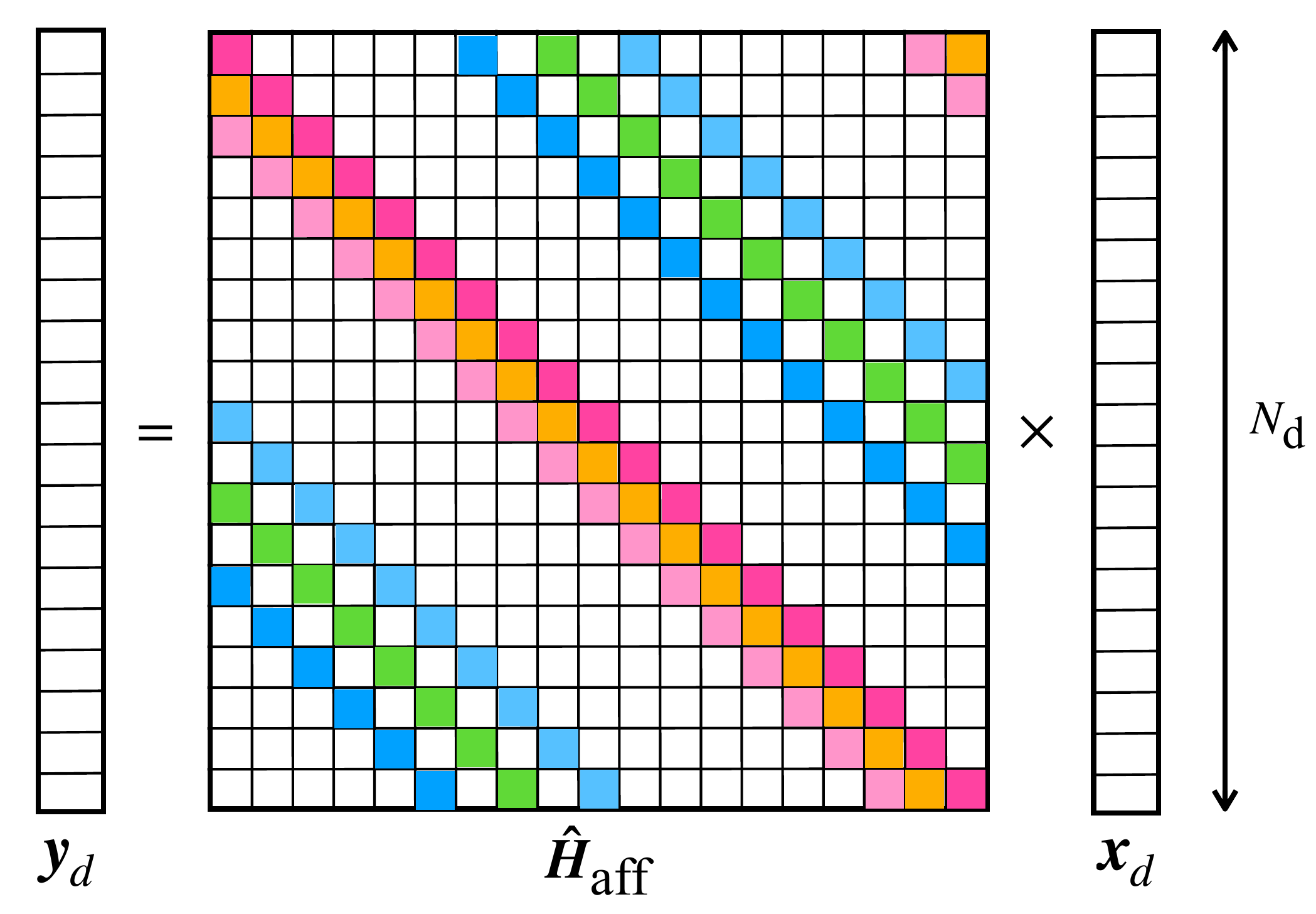}}
    \vspace*{-0.15cm} 

%\vspace{1mm}
\caption{The effective affine domain channel matrices of the proposed system (when $\chi=2$, and for a $6$-tap channel).}
\label{fig:recon}\vspace{-6mm}
\end{figure*}

%Directly transforming the affine domain symbols in \eqref{equ:affineIO} causes challenges since some of these symbols experience an additional phase term $ D'_i[m]$. To overcome the  presence $ D'_i[m]$ and to convert the effective discrete affine domain IOR of ZP-AFDM to be characterized by a circular convolution,

~~In the previous subsection, we mention that $ D'_i[m]$ represents an additional phase term.
To further understand the impact of $ D'_i[m]$, in Fig. \ref{fig:recon}(a), we visualize the affine domain channel matrix $\boldsymbol{H}_\textrm{aff}$, which is characterized by the affine domain IOR in the matrix format 
\begin{equation}
\begin{aligned}
    \boldsymbol{y} = &\boldsymbol{H}_\textrm{aff}\boldsymbol{x}+\boldsymbol{w}.%\\
    %=&\sum_{i=1}^{P}\boldsymbol{H}_i\boldsymbol{x}+\boldsymbol{\hat{w}},
\end{aligned}
\label{equ:He}
\end{equation}
In Fig. \ref{fig:recon}, we consider a $6$-path channel with the normalized delays and Doppler shifts for the six paths are $(0,0)$, $(0,-2)$, $(0,2)$, $(1,0)$, $(1,-1)$, and $(1,1)$. We set $c_1=\frac{10}{2N}$, where $\chi=2$. In these figures, each color corresponds to the effective channel of a specific path. Additionally, the section of the channel matrix outlined by the blue box experiences the extra phase term $ D'_i[m]$.

By examining \eqref{equ:Deffective} and Fig. \ref{fig:recon}(a), we observe that the first $k_{\textrm{max}}$ symbols and the last $L_2 = k_\textrm{max}+2c_1Nl_\textrm{max}$ symbols in the received affine domain symbol vector $\boldsymbol{y}$ are impacted by the additional phase term $D'_i[m]$.
Thus, %From \eqref{equ:Deffective} and Fig. \ref{fig:recon}(a), we find that 
by letting the \textit{last} $k_{\textrm{max}}$ symbols in the transmitted affine domain symbol vector $\boldsymbol{x}$ be zero, the influence that the additional phase term has on the first $k_{\textrm{max}}$ symbols in $\boldsymbol{y}$ can be mitigated. Moreover, by letting the \textit{first} $L_2$ symbols in $\boldsymbol{x}$ to be zero, the influence that the additional phase term has on the last $L_2$ symbols in $\boldsymbol{y}$ can be mitigated. 
Considering these, to mitigate the influence of the additional phase term, we let $L_\textrm{z}\triangleq L_2+k_\textrm{max}$ = $2k_\textrm{max}+2c_1Nl_\textrm{max}$ symbols in $x[m]$ to be zero in our scheme. 
With this affine domain zero-padding, the affine domain transmitted symbol $x[m]$ for $m\in\{0,1,\cdots,N-1\}$, becomes
\begin{equation}
    x[m] = \begin{cases}
        x_\textrm{d}[m-L_2],&L_2\leq m\leq N-k_\textrm{max}-1,\\
        0,&\textrm{otherwise},
    \end{cases}\!\!\!\!\!\!\!\!
    \label{equ:xm'}
\end{equation}
where $x_\textrm{d}[m'], \forall m' \in \{0,1,\dots,N_\textrm{d}-1\}$, are the information-bearing symbols, and $N_\textrm{d}=N-L_\textrm{z}$ is the total number of information-bearing symbols.

For this ZP-AFDM scheme, the IOR between the $N_\textrm{d}$ information-bearing affine domain symbols $x_\textrm{d}[m']$ and the $N$ received affine domain symbols $y[m]$ %, $\forall m\in\{0,1,\cdots,N-1\},$ 
becomes \vspace{-1mm}
\begin{align}
    y[m] =&\sum_{i\in I_i^m} \hat{h}_i x_\textrm{d}[m-\hat{l}_i]  e^{-\frac{j\pi m k_i}{c_1N^2}}+\bar{w}[m],
    \label{equ:ydm}
\end{align}
where $\hat{h}_i=h_i e^{\frac{j\pi k_i^2}{2c_1N^2}}$, $\hat{l}_i = L_2+k_i-2c_1Nl_i$, which is the effective delay in this ZP scheme, and the path index set $I_i^m$ includes the paths that satisfy $0\leq m-\hat{l}_i\leq N_\textrm{d}-1$. 
As can be observed, the impact of the additional phase term $D'_i[m]$ that exists in \eqref{equ:affineIO} disappears in \eqref{equ:ydm}. 
Also, the affine domain channel matrix $\boldsymbol{\bar{H}}_\textrm{aff}$, characterized by 
\begin{equation}
    \boldsymbol{y} = \boldsymbol{\bar{H}}_\textrm{aff}  \boldsymbol{x}_\textrm{d} + \boldsymbol{\bar{w}},
    \label{equ:He'}
\end{equation} 
has now been transformed into a banded matrix without additional phase terms, as visualized in Fig. \ref{fig:recon}(b). 
Furthermore, %when phase variations are ignored, 
the effective IOR of the ZP-AFDM in \eqref{equ:ydm} can now be viewed as an approximately linear convolution with a slowly varying phase term. 
%
%\color{blue}
Motivated by conventional OFDM systems for LTI channels, to leverage a simple one-tap frequency domain equalizer, it is necessary to have an IOR with approximately circular convolution.
Thus, to convert the IOR in \eqref{equ:ydm} into one with approximately circular convolution, we superimpose the last $L_\textrm{z}$ symbols sequence $y[m], \forall m \in\{N-L_\textrm{z},\dots, N-1\}$, onto the first $L_\textrm{z}$ symbols.%\footnote{%\color{blue}To prevent the original diagonals from overlapping with the cyclic superimposed diagonals, the condition $L_\textrm{z}<\frac{N}{2}$ must be met. In practice, since the number of subcarriers $N$ is large, this condition is generally satisfied.}.  
\begin{figure*}[b]
\vspace*{-0.3cm} 
\vspace*{-0.2cm} 
\hrulefill
    \begin{align}
        Y_\textrm{d}[k] =&X_\textrm{d}[k]\underbrace{\sum_{i=1}^{P} \hat{h}_i    e^{\frac{j2\pi k\hat{l}_i}{N_\textrm{d}}}\kappa_{N_\textrm{d},\hat{l}_i}\left(-\frac{k_i N_\textrm{d}}{2c_1N^2}\right)}_{H_\textrm{FoA}^\textrm{diag}[k]} + \underbrace{\sum_{k'=0,k \neq k'} ^{N_\textrm{d}-1}X_\textrm{d}[k']\sum_{i=1}^{P} \hat{h}_i    e^{\frac{j2\pi k'\hat{l}_i}{N_\textrm{d}}} \kappa_{N_\textrm{d},\hat{l}_i}\left(k'-k-\frac{k_i N_\textrm{d}}{2c_1N^2}\right)}_{\eta_\textrm{FoA}[k]}+\hat{W}[k].\label{equ:Ydk2}\tag{22}
    %&=\sum_{k'=0}^{N_\textrm{d}-1} H_\textrm{FoA}(k,k')X_\textrm{d}[k'],
    \end{align}
\vspace*{-5mm}
\end{figure*}
This process results in the  cyclically superimposed reconstructed received symbols $y_\textrm{d}[m]$ for $m \in \{0,1,\dots,N_\textrm{d}-1\}$ as %$\forall m\in\{0,1,\cdots, N_\textrm{d}-1\}$.
\begin{equation}
    y_\textrm{d}[m]=\begin{cases}
        y[m]+y[m+N_\textrm{d}], & 0\leq m\leq L_\textrm{z}-1,\\
        y[m], &\textrm{otherwise}.
    \end{cases}
    \label{equ:yrm1}
\end{equation}
Using \eqref{equ:affineIO} and \eqref{equ:xm'}, $y_\textrm{d}[m]$ can be expressed as 
\begin{equation}
\begin{aligned}
    \!\!\!\!y_\textrm{d}[m]=&\sum_{i=1}^{P} \hat{h}_i e^{-j2\pi \frac{k_i}{2c_1N}\frac{m+\bar{m}_i}{N}}  x_\textrm{d}\left[(m{-}\hat{l}_i)_{N_\textrm{d}}\right]
    +\hat{w}[m],\!\!\!
\end{aligned}
     \label{Equ:yrm}
\end{equation}
where $\hat{w}[m]$ represents the noise after reconstruction, and  $\bar{m}_i = N_\textrm{d}$ for $ m\geq N_\textrm{d}+k_i-2c_1Nl_i+L_2$ and $\bar{m}_i = 0$, otherwise. Expression \eqref{Equ:yrm} can be vectorized using the corresponding affine domain effective matrix $\boldsymbol{\hat{H}}_\textrm{aff}$ as 
\begin{equation}
    \boldsymbol{y_\textrm{d}} = \boldsymbol{\hat{H}}_\textrm{aff}\boldsymbol{x_\textrm{d}}+\boldsymbol{\hat{w}}.\label{equ:yrMat}
\end{equation}
Visual illustration of $\boldsymbol{\hat{H}}_\textrm{aff}$ is shown in Fig. \ref{fig:recon}(c). 
%As can be observed in Fig. \ref{fig:recon}(b), 
It has to be noted that as a result of this receiver side cyclically superimposed signal reconstruction process, the red-dotted box part in $\boldsymbol{\bar{H}}_\textrm{aff}$ in Fig. \ref{fig:recon}(b) is shifted to the top of $\boldsymbol{\bar{H}}_\textrm{aff}$, resulting in the circular structure shown in $\boldsymbol{\hat{H}}_\textrm{aff}$ in Fig. \ref{fig:recon}(c).

\subsection{Frequency of Affine Domain and One-tap Equalizer}
~~The zero-padding at the transmitter and the reconstruction operation at the receiver enable to characterize the effective discrete affine domain IOR of ZP-AFDM in \eqref{equ:yrMat} by an approximately circular convolution with a slowly varying phase term. Moreover, %as illustrated in subsection III-A, 
the effective frequency shifts experienced by the affine domain symbols $\bar{f}_{i,\textrm{aff}}=\frac{k_i}{2c_1N}=\frac{k_i}{\chi(2k_\textrm{max}+1)}$ are relatively low. Leveraging these, in order to perform equalization in a low-complex manner, we first transform the reconstructed affine domain symbols $Y_\textrm{d}[m]$ in \eqref{Equ:yrm} into a \textit{new domain by using the Fourier transform, which we refer to as the FoA domain}.
FoA domain symbols can be obtained by applying discrete Fourier transform (DFT) to $Y_\textrm{d}[m]$, yielding
\begin{align}\label{equ:Ydk}\setcounter{equation}{20}
    &Y_\textrm{d}[k] = \frac{1}{\sqrt{N_\textrm{d}}}\sum_{m = 0}^{N_\textrm{d}-1}y_\textrm{d}[m] e^{-j2\pi \frac{mk}{N_\textrm{d}}}, \forall k {\in}[0,N_\textrm{d}-1] \\
    &=\sum_{i=1}^{P}\! \hat{h}_i \!\!\sum_{k'=0}^{N_\textrm{d}-1}\!\!X_\textrm{d}[k']
 e^{\frac{-j2\pi k'\hat{l}_i}{N_\textrm{d}}}\kappa_{N_\textrm{d},\hat{l}_i}\left(k'{-}k{-}\frac{k_i N_\textrm{d}}{2c_1N^2}\right)+\hat{W}[k],\notag
%&=\sum_{k'=0}^{N_\textrm{d}-1} H_\textrm{FoA}(k,k')X_\textrm{d}[k'],
\end{align}
%for \(\forall k {\in}[0,N_\textrm{d}-1]\), 
where $X_\textrm{d}[k] = \frac{1}{N_\textrm{d}}\sum_{m = 0}^{N_\textrm{d}{-}1}x_\textrm{d}[m]e^{-j2\pi\frac{mk}{N_\textrm{d}}}$ is the FoA domain symbols corresponding to transmitted affine domain data symbols $x_\textrm{d}[m]$, and $\kappa_{N_\textrm{d},\hat{l}_i}(\phi) {=} \frac{1}{N_\textrm{d}}\sum_{m=\hat{l}_i}^{N_\textrm{d}+\hat{l}_i-1}\!e^{\frac{j2\pi m }{N_\textrm{d}}\phi}$ is a sinc-like function. 

Owing to this sinc-like nature of $\kappa_{N_\textrm{d},\hat{l}_i}(\phi)$, we can deduce that the contributions to  $Y_\textrm{d}[k]$ in \eqref{equ:Ydk} in our ZP-AFDM scheme is predominantly determined  only by a limited number of symbols $X[k']$  around the index $k'=k$. We find that $Y_\textrm{d}[k]$ in \eqref{equ:Ydk} can be decoupled based on the contribution from $X_\textrm{d}[k]$ and from $X_\textrm{d}[k']|_{k'\neq k}$ as in \eqref{equ:Ydk2}. Therein, $H_\textrm{FoA}^\textrm{diag}[k]$ is the diagonal elements in the FoA domain channel matrix $\boldsymbol{\hat{H}}_\textrm{FoA} \triangleq  \boldsymbol{F}_{N_\textrm{d}}\boldsymbol{\hat{H}}_\textrm{aff}\boldsymbol{F}_{N_\textrm{d}}^\textrm{H}$, given by $H_\textrm{FoA}^\textrm{diag}[k] = \hat{H}_\textrm{FoA}[k,k]$ with 
\begin{align}\setcounter{equation}{22}
\vspace{-0.2cm}
    \hat{H}_\textrm{FoA}[k,k'] = \sum_{i=1}^{P}\hat{h}_ie^{\frac{j2\pi k\hat{l}_i}{N_\textrm{d}}}\kappa_{N_\textrm{d},\hat{l}_i}\left(k'-k-\frac{k_i N_\textrm{d}}{2c_1N^2}\right). \label{equ:Hfoakk}
    \vspace{-0.2cm}
\end{align}
Moreover, $\eta_\textrm{FoA}[k]=\sum_{k'=0,k'\neq k}^{N_\textrm{d}-1}\hat{H}_\textrm{FoA}[k,k']X_\textrm{d}[k']$ represents the FoA domain interference experienced by the $k$-th FoA domain symbol.
We find that as $\chi$ or effectively $c_1$ increases, the term $\frac{k_i N_\textrm{d}}{2c_1N^2}$ in $\kappa_{N_\textrm{d},\hat{l}_i}\left(k'-k-\frac{k_i N_\textrm{d}}{2c_1N^2}\right)$ inside $\eta_\textrm{FoA}[k]$ approaches zero.
Given this, when we set a large $\chi$ in our proposed ZP-AFDM design, $Y_{\textrm{d}}[k]$ %in \eqref{equ:Ydk2} 
becomes predominantly impacted only by the symbol $X_\textrm{d}[k]$ and the interference $\eta_\textrm{FoA}[k]$ becomes very weak. 

\begin{figure}
    \centering
    %\vspace*{-0.4cm} 
        \includegraphics[width=0.98\linewidth]{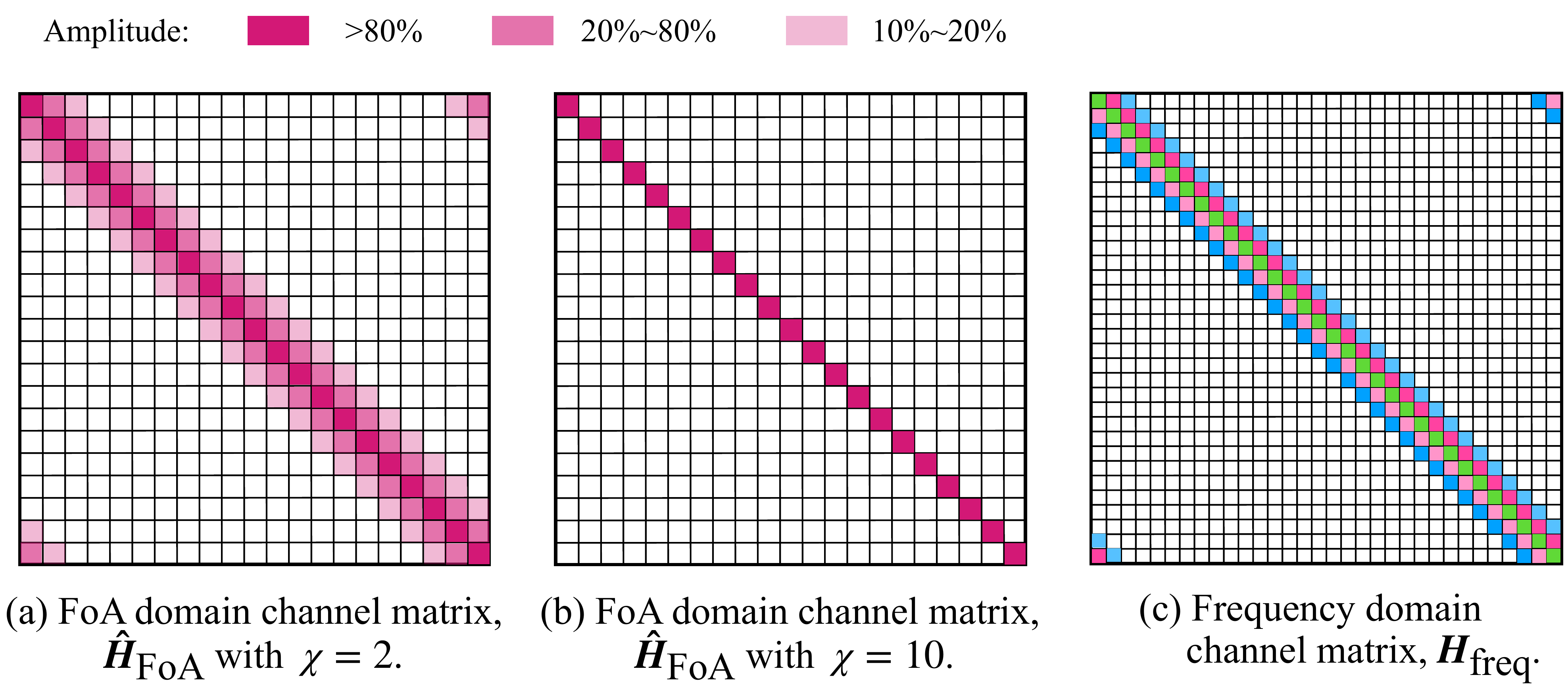}
    \vspace*{-0.3cm} 
    \caption{Effective channel matrices.}
    \label{fig:Hfoa}
\end{figure}

To further understand the FoA domain effective channel $\boldsymbol{H}_\textrm{FoA}$, we visually illustrate it in Fig. \ref{fig:Hfoa} for the parameters adopted for Fig. 3. The classical frequency domain channel matrix $\boldsymbol{H}_\textrm{freq}$ in the considered doubly selective channel is also plotted in Fig. \ref{fig:Hfoa} for comparison.  %It is interesting to observe that compared with the classical frequency domain channel matrix shown in Fig. \ref{fig:Hf}, the band of $\boldsymbol{H}_\textrm{FoA}$
It can be observed that $\boldsymbol{H}_\textrm{FoA}$ is nearly a diagonal matrix, %since the frequency shift term $e^{\frac{-j2\pi m k_i}{2c_1N^2}}$ is significantly small. 
with the width of the diagonal band determined by the function $\kappa\left(k, \frac{j2\pi k_i}{2c_1N^2}\right)$. Moreover, it is interesting to observe that compared $\boldsymbol{H}_\textrm{freq}$, the band of $\boldsymbol{H}_\textrm{FoA}$ matrix is much narrower.

%It has to be noted that when the effective frequency shift $f_{i,\textrm{aff}}=\frac{k_i}{2c_1N}$ is small, the absolute value of $\kappa\left(0, \frac{j2\pi k_i}{2c_1N^2}\right)$ approaches 1, indicating that most of the power is concentrated along the diagonal, making the diagonal matrix approximation reasonable. However, if $f_{i,\textrm{aff}}$ is larger, the approximation may degrade, as the sub-lobes contribute to noise, potentially affecting performance.

%Based on the diagonal-like structure of the effective FoA channel matrix $\boldsymbol{H}_\textrm{FoA}$ with small side-lobes, %as shown in Fig. \ref{fig:AFDMonetap}, \textit{we propose a one-tap equalizer in the FoA domain symbols $Y_\textrm{d}[k]$ in \eqref{equ:Ydk2}}.

As highlighted above, for large $\chi$ and $c_1$ in our proposed ZP-AFDM scheme,  $Y_\textrm{d}[k]$ in \eqref{equ:Ydk2} becomes predominantly impacted only by the symbol $X_\textrm{d}[k]$. Leveraging this, \textit{we propose a one-tap equalizer in the FoA domain on the symbols $Y_\textrm{d}[k]$ in \eqref{equ:Ydk2}}.
%
%Since $\left|\kappa_{N_\textrm{d},\hat{l}_i}(\phi)\right|^2$ of the power is concentrated on the diagonal, we treat only the diagonal elements as signal power and the power outside the diagonal (the sidelobes) as interference. 
In particular, we equalize $Y_\textrm{d}[k]$ using $E[k]$ as \vspace{1mm}
\begin{equation}
    \hat{X}_\textrm{d}[k]=Y_\textrm{d}[k]E[k],~~~~~~~~~ \forall k\in \{0,1,\dots,N_\textrm{d}-1 \},
    \label{equ:Equalization} \vspace{1mm}
\end{equation}
where 
$E[k] = (H_\textrm{FoA}^\textrm{diag}[k]^*)/\left({|H_\textrm{FoA}^\textrm{diag}[k]|^2+\sigma^2_\textrm{total}}\right)$.
Here, $\sigma^2_\textrm{total} = \hat{\sigma}^2+\sigma^2_\textrm{I}$ is the total noise and interference power, which includes noise $\hat{\sigma}^2 = \frac{N}{N_\textrm{d}}\sigma^2$ and the weak interference power $\sigma^2_\textrm{I}$, which can be calculated independent of $X_\textrm{d}[k]$ as 
    \begin{align}
    \sigma^2_\textrm{I} %=&\sum_{i=1}^{P}\sum_{k = 1}^{N_\textrm{d}-1}\left|h_i\kappa_{N_\textrm{d},\hat{l}_i}\left(k-\frac{k_i N_\textrm{d}}{2c_1N^2}\right)\right|^2\notag \\
        =&\sum_{i=1}^{P}|h_i|^2\left(1-\left|\kappa_{N_\textrm{d},\hat{l}_i}\left(-\frac{k_i N_\textrm{d}}{2c_1N^2}\right)\right|^2\right).
        \label{equ:interference}
    \end{align}
    \color{black}
After the one-tap equalizer in \eqref{equ:Equalization}, as shown in Fig. \ref{fig:AFDMonetap}, the equalized FoA domain symbols $\hat{X}_\textrm{d}[k]$ are sent to a $N_\textrm{d}$-point IDFT to obtain the equalized affine domain symbol
\begin{align}
    \hat{x}_\textrm{d}[m]&=\frac{1}{\sqrt{N_\textrm{d}}}\sum_{k = 0}^{N_\textrm{d}-1} \hat{X}_\textrm{d}[k]e^{j2\pi\frac{mk}{N_\textrm{d}}}\approx x_\textrm{d}[m]+\hat{w}[m],
    \label{equ:xhdm}
\end{align}
where $\hat{w}[m]$ is the noise and interference after equalization.

\begin{remark}
In this one-tap equalization process, the width of the band in the FoA domain channel matrix $\boldsymbol{\hat{H}}_\textrm{FoA}$ plays a significant role in determining the detection performance. A larger $c_1$ can be chosen to narrow the band, allowing the one-tap equalizer to more effectively compensate for channel effects.
However, increasing $c_1$ also raises the cost of zero-padding. Therefore, a trade-off between BER and spectral efficiency must be considered, depending on the communication environment and specific requirements.
\end{remark}

\section{Numerical Results}
In order to evaluate the performance of our proposed ZP-AFDM scheme with one-tap FoA domain equalizer, we conduct numerical evaluations using the EVA channel model while considering the maximum UE speed of $500$ km/h. We consider QPSK modulation, the bandwidth of $B=2$ MHz, carrier frequency $f_c=2$ GHz, and $N {=} 4096$ \cite{Shen2025TWC}. For this scenario, the maximum normalized delay is $l_\textrm{max}=5$ and the maximum normalized Doppler is $k_\textrm{max}=4$.

\begin{figure}
\vspace*{-0.4cm}
    \centering
    \includegraphics[width=0.95\linewidth]{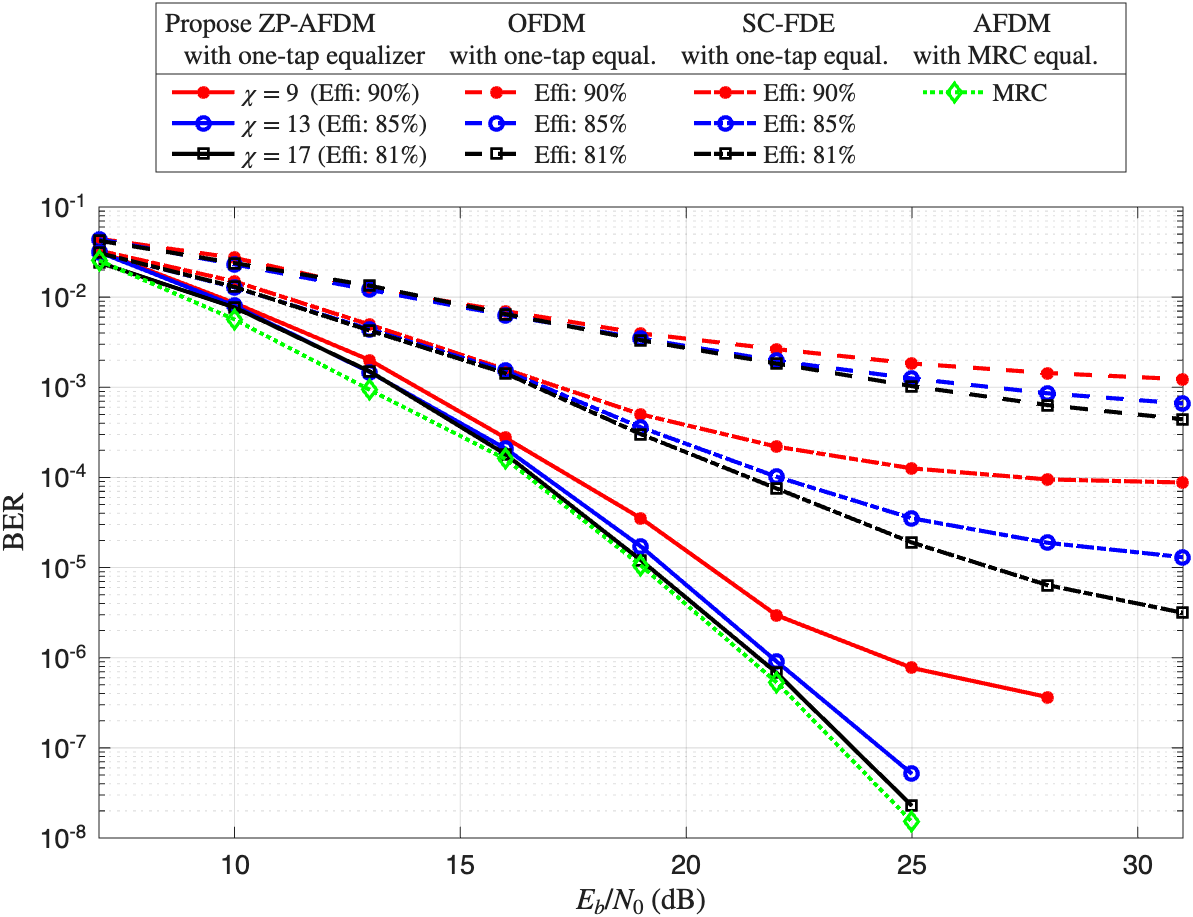}
    \vspace*{-0.3cm}
    \caption{BER vs $E_b/N_0$ for our proposed AFDM one-tap equalizer and benchmarks (EVA channel with maximum UE speed 500 km/h, QPSK).}
    \label{fig:onetapBER}
    \vspace*{-0.6cm}
\end{figure}
Fig. \ref{fig:onetapBER} illustrates the BER performance of our ZP-AFDM design versus $E_b/N_0$ for different levels of efficiency and zero-padding, which is controlled by $\chi$.
We recall that high $\chi$ leads to a higher number of zero-padding, thus lower transmission efficiency.
For comparison, the BER of the following schemes are also plotted: (i) AFDM performance with multi-tap MRC equalization as in \cite{bemani2023affine}, (ii) OFDM with a one-tap equalizer for the same CP overhead as the proposed ZP-AFDM scheme with \(\chi = 9, 13, 17\), %for example, with $\chi = 9$, the corresponding subcarrier spacing $45.5$ kHz is and the number of subcarrier is 44,
(iii) Single-Carrier frequency domain equalizer (SC-FDE) with a one-tap equalizer for the same CP overhead as the proposed ZP-AFDM scheme. For example, for the OFDM system with one-tap equalizer, $\chi = 9$ corresponds to $83$ OFDM symbols transmitted within the duration and bandwidth of interest, with CP $10\%$ overhead, %($90\%$ efficiency), 
subcarrier spacing $45.5$ kHz, and $44$ subcarriers.

As can be observed in Fig. \ref{fig:onetapBER},   with \color{black} higher values of $\chi$ (e.g., $\chi = 17$), our proposed ZP-AFDM scheme with one-tap FoA domain equalizer provides comparable performance to AFDM with a multi-tap MRC equalizer. %approaches the optimal performance of a multi-tap equalizer, 
This demonstrates that our design can significantly reduce AFDM computational complexity %(i.e., to the lowest possible setting) 
without compromising the BER performance, although a slight spectral efficiency reduction is observed. %Furthermore, its BER performance is much better than OFDM and SC-FDE due to the diversity.
As $\chi$ decreases, the BER starts to degrade, and an error floor emerges at higher SNRs. This occurs because a lower \(\chi\) increases the effective frequency shift in the FoA domain, amplifying side lobe power and degrading the BER performance. %Furthermore, Fig. \ref{fig:onetapBER2} illustrates the BER performance in the off-grid Doppler shift scenario and it is similar compared with the on-grid scenario.
We highlight that even at low \(\chi\) values, on one hand, our design significantly outperforms OFDM and SC-FDE with a one-tap equalizer. On the other hand, with $\chi = 9$, the resulting error floors only occur at BER values of ${10}^{-6}$ or lower, which is below typical acceptable BER levels  for uncoded systems \color{black} \cite{andrews2014will}. 

\begin{figure}[t]
%\vspace*{-0.5cm}
    \centering
    \includegraphics[width=0.8\linewidth]{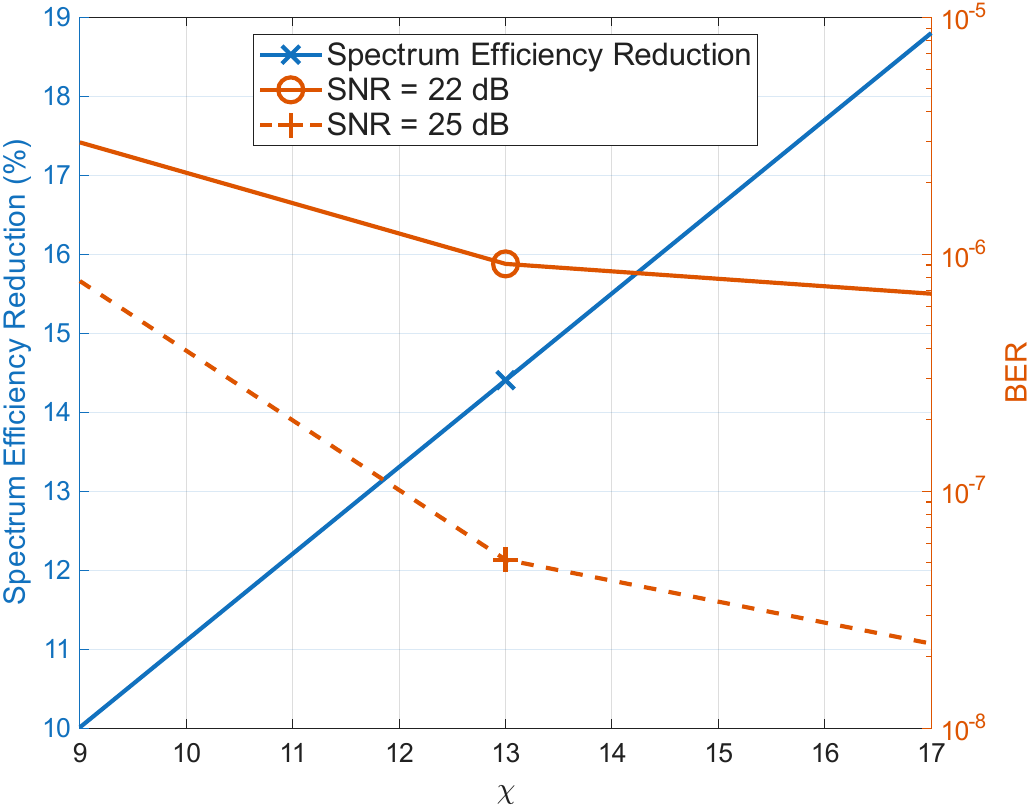}
    \vspace*{-0.3cm}
    \caption{Efficiency of ZP-AFDM with different $\chi$.}
    \label{fig:efficiency}
    \vspace*{-0.4cm}
\end{figure}

As shown in Fig. \ref{fig:onetapBER}, although increasing $c_1$ improves the BER performance of our design, it comes at a cost. To illustrate this, in Fig. \ref{fig:efficiency}, we plot the spectrum efficiency reduction of the proposed ZP-AFDM and the BER curves for various $\chi$. We observe that as \(\chi\) increases, the BER decreases; however, this also reduces the spectral efficiency of the proposed ZP-AFDM system. Furthermore, beyond a certain \(\chi\) value, the BER improvement plateaus, likely when interference becomes significantly smaller than noise power. These suggest that an optimal \(\chi\) value must be carefully selected to balance minimized BER with efficient resource use.
For instance, as shown in Fig. \ref{fig:efficiency}, if the system SNR is $22$ dB, setting \(\chi\) to $13$ achieves a good performance trade-off, while at an SNR of $25$ dB, \(\chi = 17\) is needed to obtain a good trade-off.

\section{Conclusion}
In this paper, we proposed a one-tap equalizer for ZP-AFDM scheme that effectively simplifies the equalization process for doubly selective channels. 
By carefully configuring AFDM parameters $c_1$ and $c_2$, performing zero-padding and reconstructing operation, the simplified IOR is obtained and it is subsequently transferred to the FoA domain by DFT. In the FoA domain, the effective channel matrix is nearly diagonal.
This property enables the proposed equalizer to achieve both good performance and low complexity, resulting in high resilience, particularly in high-speed conditions.
Simulation results demonstrate that the equalizer maintains strong performance with full diversity and limited redundancy, positioning ZP-AFDM as a viable approach for future wireless communication systems. %This work advances the practical applicability of AFDM and sets the foundations for its integration into next-generation high-mobility communication systems.

\end{document}